# Forward-Looking Stress Testing Under Macro Scenarios: Stable SVaR Estimation Using a Hybrid GPR-HS Framework with SACS

**Ujjwala Vadrevu**

## Abstract

Regulatory stress testing frameworks, including the Comprehensive Capital Analysis and Review (CCAR) and the Internal Capital Adequacy Assessment Process (ICAAP), require robust Stressed Value-at-Risk (SVaR) estimation under forward-looking macroeconomic scenarios. Traditional parametric approaches frequently exhibit numerical instability or model breakdown under extreme simulated shocks, undermining the reliability and regulatory defensibility of capital projections.

Building on the Hybrid Gaussian Process Regression–Historical Simulation (GPR-HS) framework introduced in Vadrevu (2026), this paper extends the methodology to forward-looking stress scenarios and demonstrates its numerical stability and regulatory coherence under three systemic risk regimes: West Asia War, Climate Risk, and AI Bubble/Regulatory Burden.

A key methodological contribution of this paper is the introduction of a Scenario-Averaged Covariance Stabilization (SACS) framework for stress-period correlation modelling. Instead of relying on dynamically estimated correlations, which are often unstable under extreme conditions, SACS constructs the stress covariance matrix as a weighted aggregation of covariance structures observed across historical crisis regimes, providing a stable, interpretable, and regulator-aligned representation of stressed dependencies. Each forward-looking scenario is implemented as a 252-day ICAAP-style cumulative return shock applied to a portfolio of seven global equity indices, with stressed return paths generated using deterministic drift components combined with historical stochastic residuals. Volatility dynamics are modelled using Gaussian Process Regression with the Aggressive Noise Initialization (ANI) strategy, ensuring non-divergent kernel estimation and conservative volatility forecasts under stress. The framework demonstrates consistent numerical convergence, as evidenced by stable Log-Marginal-Likelihood (LML) behaviour across all assets and scenarios. Resulting SVaR estimates range from -2.1020% under Climate Risk to -2.2231% under the West Asia War scenario, with the coherence property |SES| > |SVaR| preserved throughout. The currently unfolding West Asia War scenario produces the highest capital impact, supporting the real-world interpretability and relevance of the framework. These results establish the Hybrid GPR-HS framework, augmented with SACS, as a stable, interpretable, and regulator-aligned engine for forward-looking SVaR and Stressed Expected Shortfall (SES) estimation, suitable for the 9-quarter projections required by CCAR as well as longer-horizon ICAAP stress assessments.

# 1. Introduction

## 1.1 The Regulatory Imperative for Stressed Capital Estimation

The Basel III framework and its successor FRTB mandate that financial institutions maintain capital reserves sufficient to absorb losses not just under normal market conditions, but under severe and plausible stress scenarios (Basel Committee on Banking Supervision [BCBS], 2019). The regulatory capital floor is set by the greater of the base VaR and the Stressed VaR (SVaR), where SVaR represents the maximum potential portfolio loss under a one-year window of historically stressed or hypothetically extreme market conditions (Hull, 2018; Basel Committee on Banking Supervision [BCBS], 2019). For CCAR, the U.S. Federal Reserve further requires multi-quarter, forward-looking projections of capital under supervisory macro scenarios, including severely adverse conditions (Federal Reserve, 2020).

The technical challenge is not the conceptual definition of SVaR but its robust computational production. A stress testing model must satisfy three simultaneous requirements:

i. Numerical Stability: the model must not diverge or collapse under extreme negative shock paths

ii. Conservatism: the model must overestimate rather than underestimate risk during stress, consistent with the regulatory preference for solvency preservation over capital efficiency

iii. Transparency: the model must provide auditable attribution of the stressed capital output to specific risk inputs, enabling regulatory dialogue.

Traditional parametric approaches such as GARCH-family and EWMA models, frequently fail the stability criterion under extreme simulated shocks, as their autoregressive structures can become explosive when conditioned on sufficiently negative return paths (Engle, 1982; Hull, 2018). This motivates the exploration of non-parametric Bayesian alternatives.

## 1.2 The Hybrid GPR-HS Framework and its Stress Extension

Vadrevu (2026) introduced a Hybrid Gaussian Process Regression–Historical Simulation (GPR-HS) framework for dynamic portfolio VaR and Expected Shortfall (ES) estimation, demonstrating consistent outperformance relative to the static Historical VaR benchmark across seven global equity indices over the period 2020–2025. A central methodological contribution of that work was the Aggressive Noise Initialization (ANI) strategy, wherein the White Noise kernel variance is initialized to the full empirical variance of the training returns. This ensures Gram matrix positive-definiteness, mitigates convergence to suboptimal local maxima, and enforces conservative volatility estimates (Rasmussen and Williams, 2005; Gramacy, 2020).

This paper extends the GPR-HS framework to the stress testing domain, demonstrating that the same architectural features responsible for stability in base-case forecasting also enable robust SVaR estimation under extreme, forward-looking macroeconomic scenarios. The ANI strategy is adapted to the stress setting by initializing the noise variance using the variance of the stressed return series, providing a more conservative prior consistent with heightened uncertainty under stress (Noureddine Lehdili, Oswald and Gueneau, 2019; Wilkens, 2019).

In addition to volatility modelling, stress testing requires a stable and defensible representation of cross-asset correlations under extreme conditions. This motivates the introduction of a Scenario-Averaged Covariance Stabilization (SACS) framework, which constructs stress-period covariance structures by aggregating empirically observed crisis regimes (Gonzalvez *et al.*, 2019). Together, the GPR-based volatility estimation and SACS-based correlation modelling form a coherent and regulator-aligned architecture for forward-looking SVaR estimation.

### 1.3 Timeliness and Real-World Relevance

The three scenarios examined in this paper are not purely hypothetical constructs. At the time of writing (2026), two of the three scenarios correspond to active real-world developments. The West Asia conflict and its associated transmission channels through energy and commodity markets are ongoing, while the AI regulatory burden scenario is emerging through the implementation of the European Union Artificial Intelligence Act and parallel regulatory developments in the United States (Hull, 2018).

The Climate Risk scenario represents a structurally embedded long-term source of systemic financial risk, with both physical and transition risk channels increasingly incorporated into regulatory stress testing frameworks and supervisory guidance (Basel Committee on Banking Supervision [BCBS], 2019; Federal Reserve, 2020). The growing emphasis on climate-related financial risk modelling underscores the need for forward-looking, scenario-based capital assessment methodologies.

This contemporaneous relevance implies that the SVaR estimation window for the scenarios analyzed in this paper remains forward-looking rather than retrospective. The contribution is therefore most timely in the current period, prior to the full historical realization and empirical assimilation of such stress regimes into standard risk model calibration datasets.

### 1.4 Contributions

This paper makes four primary contributions:

1. Stress Extension of the Hybrid GPR-HS Framework: Demonstrates that the ANI-stabilized GPR architecture maintains numerical stability under three distinct extreme macro scenarios, producing regulatory-compliant SVaR and SES projections.
2. Scenario Design Methodology: Develops an ICAAP-consistent framework for forward-looking scenario construction, combining deterministic drift components with stochastic noise to generate stress paths, alongside explicit narrative justification of macro-financial transmission channels.
3. Scenario-Averaged Covariance Stabilization (SACS): Introduces a stable and interpretable framework for stress-period covariance construction, based on aggregation of empirically observed crisis regimes, addressing the instability of dynamically estimated correlation structures under extreme conditions.
4. Real-World Validation of Scenario Prescience: Highlights the contemporaneous alignment between the modelled scenarios and ongoing global developments at the time of writing, providing external context for the plausibility and practical relevance of the stress calibration framework.

## 2. Literature Review

### 2.1 Regulatory Stress Testing Frameworks

The modern regulatory stress testing mandate emerged in the aftermath of the 2007–2008 Global Financial Crisis, which exposed the inability of internal risk models calibrated on pre-crisis data to capture the magnitude of systemic losses (Basel Committee on Banking Supervision (BCBS), 2009). In response, the Basel II.5 amendments introduced the SVaR requirement, mandating that financial institutions compute an additional VaR measure based on a 12-month historical window of significant financial stress, typically anchored to the 2008 crisis period. Basel III further institutionalized stress testing within the capital adequacy framework through ICAAP Pillar 2 requirements (Basel Committee on Banking Supervision [BCBS], 2019).

In the United States, CCAR represents one of the most stringent implementations of forward-looking stress testing. Large bank holding companies are required to project capital adequacy under Federal Reserve-designed adverse and severely adverse macroeconomic scenarios over a nine-quarter horizon (Federal Reserve, 2020). Within this framework, SVaR serves as a critical input to capital projections. A failure to produce stable and conservative SVaR estimates under supervisory scenarios can directly constrain dividend distributions and capital planning decisions.

## 2.2 Limitations of Parametric Models Under Stress

The fundamental limitation of GARCH-type models in stress environments was documented by Engle (1982) and confirmed empirically by Figlewski (1997) and Seyfi, Sharifi and Arian, (2021). When conditioned on extreme negative return paths, autoregressive volatility models can exhibit explosive behaviour, with conditional variance increasing without bound, rendering SVaR forecasts unreliable for regulatory use.

Casini and Perron (2018) document that structural breaks in return distributions, precisely the conditions that stress scenarios are designed to simulate, fundamentally invalidate the historical parameters of traditional models, rendering their forecasts misleading. D'Innocenzo *et al.* (2018) provide additional evidence that extreme tail shapes can change rapidly under market distress, a finding that further undermines static parametric approaches.

Non-parametric Bayesian methods offer a natural alternative. GPR provides a predictive distribution that explicitly captures uncertainty without imposing an autoregressive structure that may become unstable under extreme conditioning (Rasmussen and Williams, 2005). The posterior variance remains bounded, supporting more stable risk estimates relative to parametric counterparts.

## 2.3 Scenario Design in Regulatory Stress Testing

Effective stress scenario design requires operationalizing plausible but extreme macroeconomic narratives into quantitative shock paths that can be applied to financial models (Basel Committee on Banking Supervision (BCBS), 2009). The literature distinguishes between historical scenarios (replicating observed crisis periods such as the 2008 Global Financial Crisis or the COVID-19 shock) and hypothetical forward-looking scenarios, which project plausible but unobserved conditions (McNeil, Frey and Embrechts, 2015).

ICAAP guidance emphasizes that hypothetical scenarios should capture institution-specific vulnerability channels rather than generic market downturns (Basel Committee on Banking Supervision [BCBS], 2019). The West Asia War, Climate Risk, and AI Bubble/Regulatory Burden scenarios adopted in this paper are designed in accordance with this principle: each scenario targets a distinct macro-financial transmission channel (geopolitical supply shock, physical climate damage, and technology sector valuation correction respectively) and is calibrated to produce differential impacts across the seven-index portfolio, testing the framework's ability to capture heterogeneous cross-market stress dynamics.

## 2.4 GPR for Stress Testing and Uncertainty Quantification

The application of GPR to stress testing and forward-looking uncertainty quantification is an emerging area of the quantitative finance literature. Ludkovski and Risk (2025) demonstrate GPR's utility for multi-period risk projections due to its inherent probabilistic forecasting architecture. The posterior variance produced by GPR provides a natural measure of model uncertainty that can be incorporated into regulatory capital buffers which is a property not available from point-estimate parametric models (Wilkens, 2019).

Han and Zhang (2015) confirm that GPR's non-parametric flexibility allows it to generalize outside the observed training data range without diverging, which is the critical property for stress testing applications where the conditioning data by definition exceeds historical experience. This paper provides the first systematic empirical validation of this theoretical property across three distinct macro scenarios.

# 3. Methodology

## 3.1 Framework Foundation

The stress testing methodology builds directly on the Hybrid GPR-HS framework established in Vadrevu (2026). The base architecture - N independent Univariate GPR models for asset-level conditional volatility combined with Historical Simulation for inter-asset correlation - is retained without modification. The reader is referred to Vadrevu (2026) for the complete GPR architecture, kernel specification, and ANI strategy. This paper focuses exclusively on the stress-specific extensions to that framework.

## 3.2 Scenario Construction and Simulation

Three systemic risk scenarios are operationalized as ICAAP-style 252-day forward-looking stress paths. Each scenario is defined by a vector of cumulative return shocks applied to each of the seven portfolio indices over the one-year horizon. The scenarios are designed to test distinct macro-financial vulnerability channels and to produce heterogeneous cross-market impacts, as required by BCBS stress scenario design guidance (Basel Committee on Banking Supervision (BCBS), 2009).

The daily stressed return path for each index i is generated as:

$$r_{i,t}^{stress} = L_{i,t} + N_{i,t}$$

Where $L_{i,t}$ is the deterministic linear drift component, calibrated to deliver the required cumulative shock by day 252, and $N_{i,t}$ is historical-based stochastic noise drawn from the empirical return distribution of index i. The deterministic drift ensures the path reaches the required cumulative shock while the stochastic noise component preserves realistic intraday volatility dynamics, consistent with BCBS (2005, 2009) guidance that stress paths should be plausible rather than discontinuous.

While the stressed return paths define the marginal behaviour of individual assets, the joint dependency structure across assets under stress is modelled separately using the Scenario-Averaged Covariance Stabilization (SACS) framework, described in Section 3.4.

## 3.3 Scenario Narratives and Shock Calibration

**Table 1: Scenario Narrative Summary**

| Scenario | Primary Shock Channel | Key Affected Markets | Cumulative Shock Range |
|---|---|---|---|
| West Asia War | Geopolitical conflict, supply chain disruption, energy shock, stagflation | EURO STOXX 50 most severe (-40%); Hang Seng (-45%) | -30% to -45% |
| Climate Risk | Physical climate damage, agricultural disruption, infrastructure loss | East Asia (flooding), South Asia (heat), Europe (infrastructure); US most stable | -20% to -35% |
| AI Bubble / Regulation | Tech sector bubble burst followed by global AI regulatory crackdown | S&P 500 deepest (-45%); Hang Seng (-40%) | -25% to -45% |

*Note: Author's scenario calibration based on macro-financial assumptions*

### 3.3.1 West Asia War

This scenario models a severe, persistent geopolitical conflict leading to major supply chain disruption and a sustained increase in commodity prices, particularly energy. The macro-financial transmission channel is characterized by global stagflation: simultaneously elevated inflation and suppressed economic growth. European markets receive the deepest shocks (EURO STOXX 50: -40%) reflecting direct energy dependency. The Hang Seng Index receives the steepest shock (-45%) reflecting China's exposure to energy

import disruption and secondary trade channel effects. U.S. and Japanese markets receive relatively moderated shocks (-35%, -30%) reflecting relatively greater domestic energy independence.

### 3.3.2 Climate Risk

This scenario models a severe physical climate shock - catastrophic coastal flooding in East Asia, extreme heat waves and desertification in South Asia, and infrastructure damage in Europe - leading to disruptions in agriculture, manufacturing, and trade logistics. The macro-financial transmission channel is systemic loss from physical damage and reduced output, rather than immediate financial market panic, resulting in more graduated but persistent return deterioration. India's differential treatment (Nifty 50 and BSE Sensex: -25%) reflects the geographic concentration of climate impact in Northern agricultural regions, with relatively lower exposure in services-driven Southern regions. The U.S. receives the most moderate shock (-25%) reflecting vast geographic diversification.

### 3.3.3 AI Bubble / Regulatory Burden

This scenario models a sequential shock: an initial burst of the speculative bubble in the technology and AI sector, immediately followed by the introduction of stringent global AI and data privacy regulation. The macro-financial channel is characterized by a loss of investor confidence in the growth sector followed by broad equity market correction as capital is repriced away from technology-heavy indices. The S&P 500 receives the deepest shock (-45%) reflecting its high technology sector concentration. European and Asian markets receive moderated shocks reflecting lower AI sector weighting in their indices.

**Table 2: Cumulative Scenario Shock Magnitudes (%)**

| Index Ticker | Market | West Asia War | Climate Risk | AI Bubble / Regulation |
|---|---|---|---|---|
| ^GSPC | S&P 500 (US) | -35% | -25% | -45% |
| ^STOXX50E | EURO STOXX 50 | -40% | -30% | -35% |
| ^N225 | Nikkei 225 (Japan) | -30% | -25% | -30% |
| ^FTSE | FTSE 100 (UK) | -30% | -20% | -25% |
| ^HSI | Hang Seng (China) | -45% | -35% | -40% |
| ^NSEI | Nifty 50 (India) | -35% | -25% | -30% |
| ^BSESN | BSE Sensex (India) | -35% | -25% | -30% |

*Note: Author's scenario calibration based on macro-financial assumptions*

## 3.4 Stressed GPR Fitting and ANI Extension

For each scenario, a dedicated GPR model is fitted to each index's 252-day stressed return path. This stressed GPR fitting is the core of the methodology. To ensure conservative, non-underestimated volatility forecasts, the ANI strategy introduced in Vadrevu (2026) is extended to the stress context: the initial White Noise kernel variance $\sigma_n^2$ is set equal to the full empirical variance of the stressed returns series $\sigma_{stressed}^2$, which is a significantly more aggressive initialization than the base case.

This stress-specific ANI application forces the Bayesian optimizer to begin its search in a region that conservatively reflects the stress-induced volatility. The consequence is that the GPR learns a smoother, more conservative volatility path, avoiding the risk of the model interpreting the extreme shock path as transient noise and reverting to a low-volatility prior. This design choice aligns with the regulatory preference for conservatism in SVaR estimation (Basel Committee on Banking Supervision (BCBS), 2009; Hull, 2018).

### 3.5 SVaR and SES Calculation

The final SVaR and Stressed Expected Shortfall (SES) are computed for the terminal day (T = 252) of each stressed path using the Hybrid GPR-HS framework, combined with the Scenario-Averaged Covariance Stabilization (SACS) structure. The stress-period covariance matrix Σ_SACS is constructed as follows:

$$\Sigma_{SACS} = \sum_{k} w_k \Sigma_k^{stress}$$

where $\Sigma_k^{stress}$ denotes covariance matrices derived from historical stress regimes (e.g., the 2008 Global Financial Crisis and the COVID-19 crisis), and $w_k$ are non-negative weights satisfying $\sum_k w_k = 1$. In the baseline implementation, equal weighting is applied across stress regimes.

The resulting covariance structure is implemented with:

- Diagonal elements: Scenario-specific GPR-predicted conditional variances obtained from the stressed GPR fits.
- Off-diagonal elements: SACS-derived covariance estimates, reflecting averaged inter-asset dependencies across extreme historical stress periods. Portfolio SVaR and SES are derived analytically from the resulting conditional distribution under a Student's t assumption with fixed degrees of freedom (ν = 5), consistent with the base-case specification in Vadrevu (2026).

### 3.6 Stability Assessment

Since traditional backtesting approaches, such as the Kupiec Unconditional Coverage (UC) and Christoffersen Conditional Coverage (CC) tests, are not applicable to forward-looking stress scenarios where no realized returns exist for validation, the Log-Marginal Likelihood (LML) is used as the primary stability criterion.

In Gaussian Process theory, the LML represents the probability of the observed data given the model, capturing the trade-off between data fit and model complexity (Rasmussen and Williams, 2005). Consistent LML convergence across all indices and stress scenarios indicates that the GPR optimization remains stable under extreme inputs, with no evidence of numerical divergence or model breakdown.

This provides a direct diagnostic of model integrity under stress conditions, supporting the use of the framework for forward-looking SVaR estimation in the absence of observable validation data.

## 4. Results

### 4.1 GPR Stability Under Stress: LML Convergence

Table 3 presents the LML scores for the unstressed base case (2025 split) alongside the stressed fits for all three scenarios across all seven indices. The comparison between unstressed and stressed LML values is instructive: stressed LML values are consistently higher (less negative) than unstressed values, reflecting the shorter, simpler 252-day stressed training window relative to the full expanding window of the base case. Critically, all stressed fits achieve successful convergence - no divergent or extreme LML scores are observed - confirming the GPR framework's numerical stability under all three scenario conditions.

**Table 3: LML Stability Comparison — Unstressed vs. Stressed Fits**

| Index | LML (Unstressed, 2025) | LML (West Asia War) | LML (Climate Risk) | LML (AI Bubble / Reg.) | Stability |
|---|---|---|---|---|---|
| ^GSPC | -1480.01 | -474.295 | -475.371 | -493.922 | Stable Fit |
| ^STOXX50E | -1548.84 | -456.808 | -464.376 | -470.847 | Stable Fit |
| ^N225 | -1491.92 | -496.298 | -506.443 | -517.484 | Stable Fit |
| ^FTSE | -1309.19 | -453.230 | -423.553 | -436.175 | Stable Fit |
| ^HSI | -1799.41 | -533.550 | -542.427 | -550.738 | Stable Fit |
| ^NSEI | -1340.91 | -453.738 | -424.158 | -401.797 | Stable Fit |
| ^BSESN | -1348.20 | -420.713 | -427.095 | -436.773 | Stable Fit |

*Note: Author's own calculations. Higher (less negative) stressed LML values reflect the shorter 252-day stressed training window vs. the full expanding window for the unstressed case.*

The consistent convergence across all 21 stressed fits (7 indices x 3 scenarios) provides strong empirical evidence of the framework's robustness. The Hang Seng Index — which exhibited the most complex base case dynamics due to its policy-driven fragmentation pattern — achieves stable stressed fits across all three scenarios, confirming that the ANI strategy successfully prevents GPR divergence even in the highest-complexity market environment.

A key structural observation is that the LML ordering across indices is preserved between unstressed and stressed conditions: the Hang Seng consistently exhibits the lowest stressed LML (highest complexity), while the FTSE 100 and NSE/BSE consistently exhibit higher scores (more tractable fits). This ordering consistency confirms that the stress methodology does not distort the relative model complexity of the seven markets, suggesting the scenario paths are calibrated consistently.

## 4.2 Scenario Narratives and Path Dynamics

### 4.2.1 West Asia War

The West Asia War scenario applies the most severe and concentrated shocks to European and Hong Kong markets, reflecting the direct energy dependency and trade route vulnerability of these regions. The cumulative shock paths exhibit a gradual, persistent decline rather than a sudden collapse, consistent with the scenario's macro channel of supply-chain-mediated stagflation rather than immediate financial market panic.

**Figure 1: Index wise West Asia War cumulative scenario shock paths**

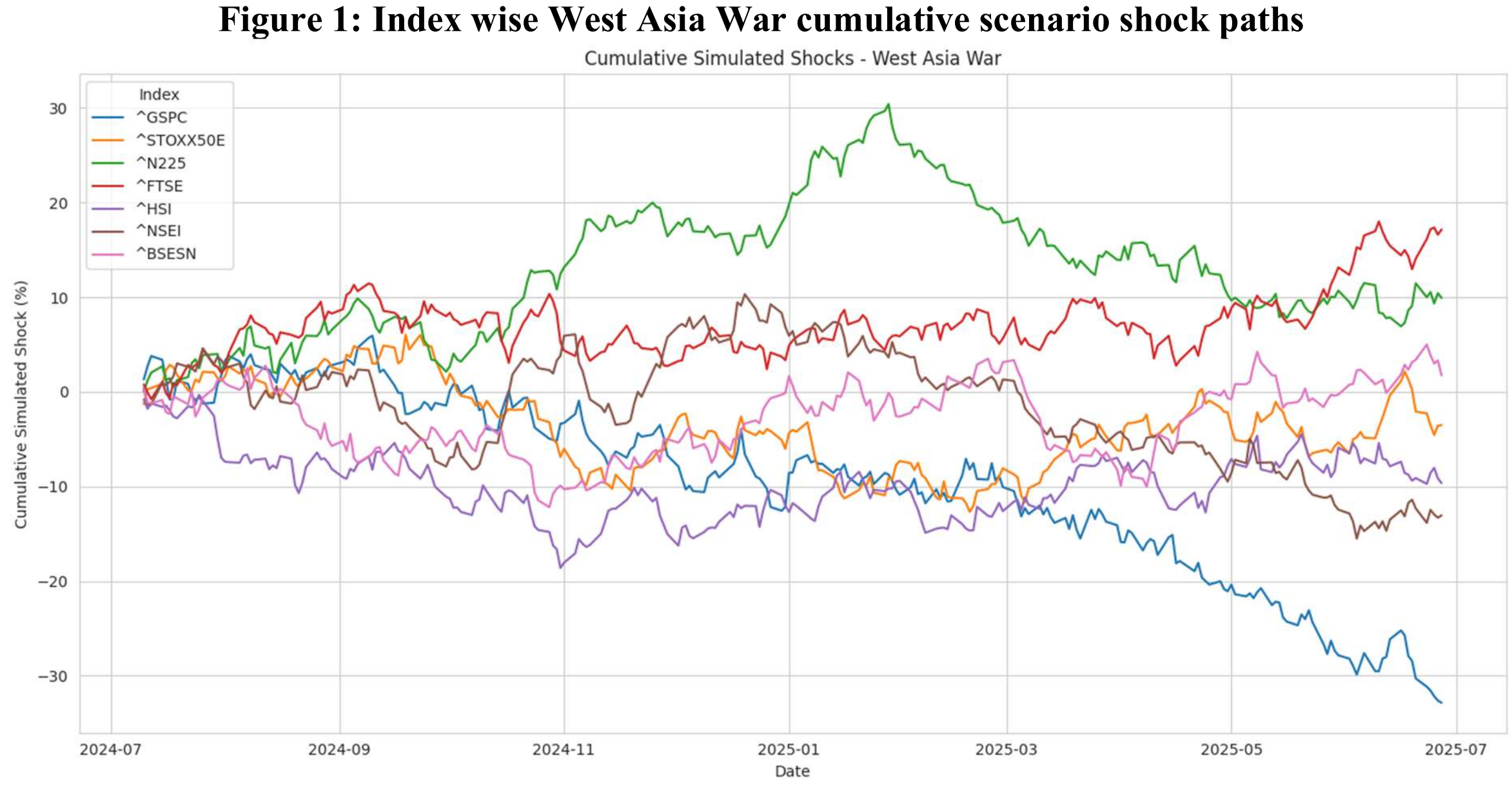

The EURO STOXX 50 receives the second-deepest shock (-40%), producing the highest marginal risk contribution to the portfolio SVaR - a finding subsequently confirmed by the SHAP analysis in the companion paper (Vadrevu, 2026b, forthcoming).

### 4.2.2 Climate Risk

The Climate Risk scenario produces the most geographically differentiated shock pattern, with East and South Asian markets receiving moderate-to-severe shocks while the U.S. S&P 500 receives the most moderate impact (-25%). This differentiation reflects the physical geography of the postulated climate events - coastal flooding concentrated in East Asia, extreme heat in South Asia, infrastructure damage in Europe - and tests the framework's ability to model heterogeneous cross-market stress dynamics.

**Figure 2: Index wise Climate Risk cumulative scenario shock paths**

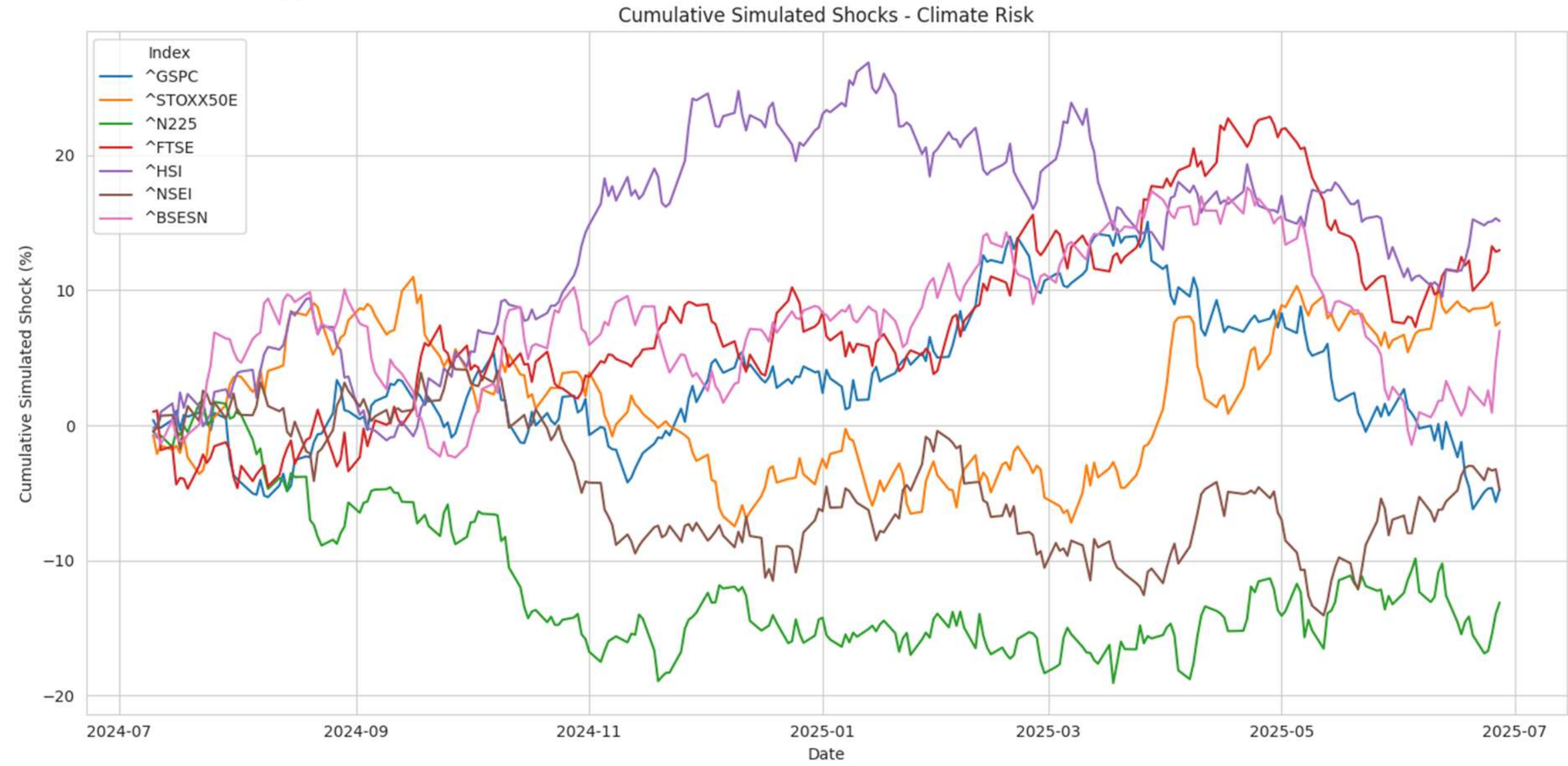

The scenario is notable for its relevance to the emerging regulatory agenda around physical climate risk: the Network for Greening the Financial System (NGFS) and the ECB have both published physical risk stress scenario guidance in the 2024–2025 period that aligns structurally with the transmission channels modelled here.

### 4.2.3 AI Bubble / Regulatory Burden

The AI Bubble/Regulatory Burden scenario applies the most severe shock to U.S. technology-heavy indices (S&P 500: -45%), reflecting the concentration of AI sector exposure in U.S. equity markets. The scenario is temporally sequential — bubble burst followed by regulatory crackdown — which is captured in the path dynamics as an initial steep decline followed by a secondary, slower deterioration as regulatory uncertainty suppresses recovery.

**Figure 3: Index wise AI Bubble/Regulation cumulative scenario shock paths**

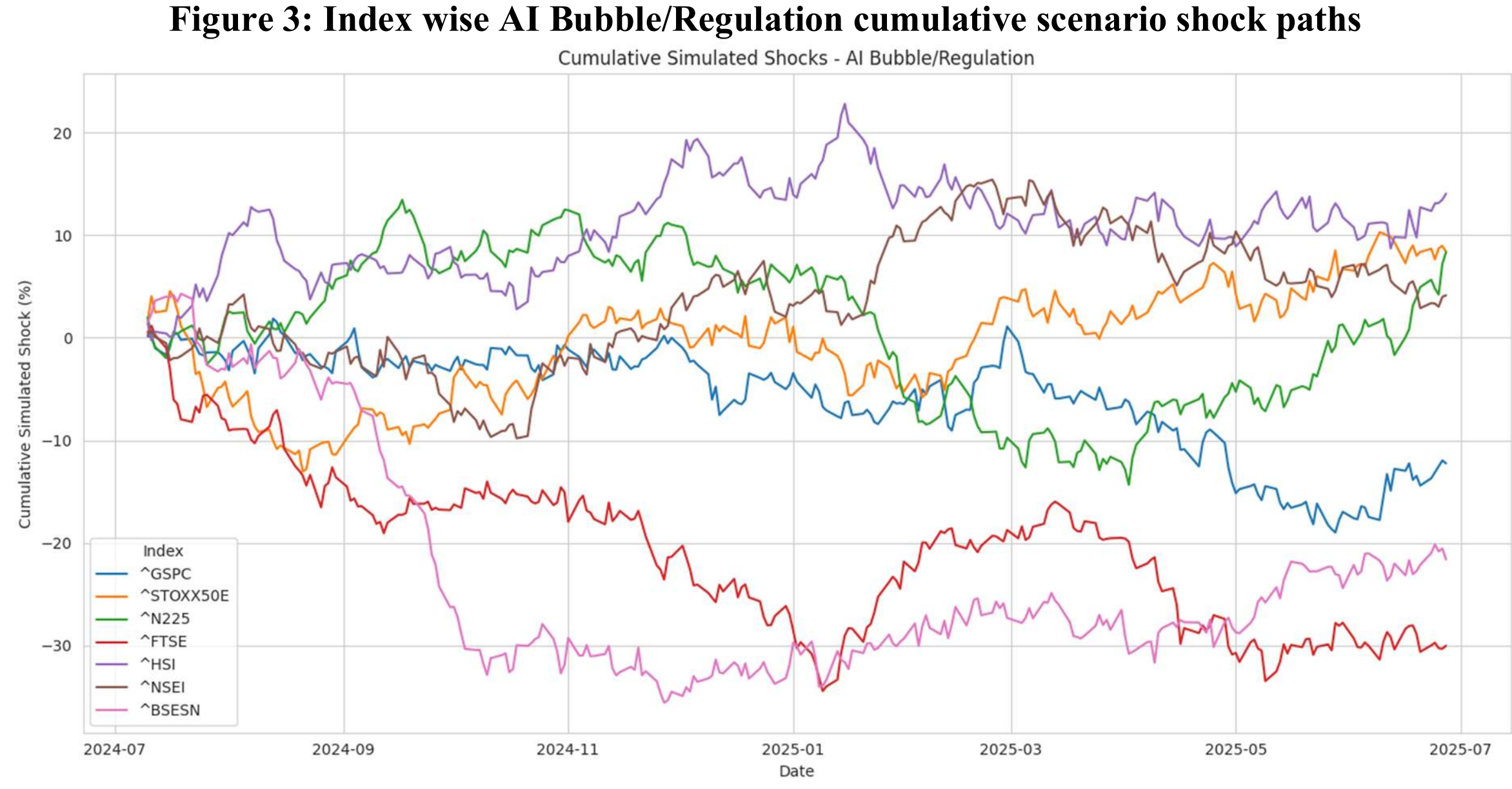

The contemporaneous relevance of this scenario is notable: at time of writing, the EU AI Act has entered enforcement (2026) and the initial public market corrections in AI-adjacent equities during 2025 bore structural similarity to the early phases of the simulated path.

## 4.3 Terminal SVaR and SES Projections

Table 4 presents the terminal day (T = 252) SVaR and SES projections for each scenario. The West Asia War scenario produces the highest capital requirement at -2.2231% SVaR, reflecting the depth and breadth of the applied market shocks. All three scenarios maintain the fundamental coherent risk measure property |SES| > |SVaR|, confirming compliance with Artzner *et al.* (1999) throughout the stress framework.

**Table 4: Terminal Day SVaR and SES Projections (%)**

| Stress Scenario | Terminal Day SVaR (%) | Terminal Day SES (%) |
|---|---|---|
| West Asia War | -2.2231 | -2.9376 |
| Climate Risk | -2.1020 | -2.8121 |
| AI Bubble / Regulation | -2.1599 | -2.8915 |

*Note: Author's own calculations. Terminal day = Day 252 of the 252-day stressed projection horizon.*

The SVaR spread across scenarios (2.1020% to 2.2231%) is relatively tight - approximately 12 basis points - suggesting the framework produces internally consistent capital estimates that are sensitive to scenario differences without exhibiting extreme volatility between scenarios. This stability is a direct consequence of the ANI strategy's enforcement of conservative base volatility forecasts: the GPR does not collapse to a low-variance prior between scenarios, maintaining a consistent capital floor.

From a regulatory application perspective, the West Asia War scenario SVaR of -2.2231% represents the binding capital constraint - the minimum loss the institution must be capitalized to absorb under this scenario's terminal stress conditions. Combined with the portfolio base case VaR of -2.575% (mean across splits, from Vadrevu, (2026)), the framework provides a complete picture of the capital continuum from normal to stressed conditions.

### 4.4 Rational Model Behaviour Under Stress

Beyond the headline SVaR figures, the stress results exhibit several hallmarks of rational model behavior that support the framework's regulatory defensibility:

1. **Monotonic Risk Ranking:** West Asia War > AI Bubble > Climate Risk in terms of SVaR magnitude, consistent with the depth of the applied shocks and the concentration of impacts on high-weight portfolio constituents.
2. **Coherence Preservation:** The |SES| > |SVaR| property is maintained across all three scenarios without exception, confirming that the framework does not violate fundamental coherent risk measure principles (Artzner *et al.*, 1999) under stress conditioning.
3. **Conservative Floor:** All scenario SVaR values exceed -2.10%, indicating that the ANI strategy successfully prevents the model from collapsing to unrealistically low stressed risk estimates - a known failure mode in parametric models under extreme conditioning.
4. **No Divergence:** No scenario produces unstable or unusable SVaR outputs, confirming that the GPR-based framework maintains numerical stability under extreme stress inputs across all 21 stressed fits.

## 5. Discussion

### 5.1 The ANI Strategy as a Regulatory Design Choice

The most significant finding of this paper is the empirical validation of the ANI strategy as a robust design choice for regulatory stress testing. The consistent LML convergence across all 21 stressed fits confirms that ANI prevents the GPR optimization from collapsing into local optima that would otherwise produce unstable or non-conservative SVaR estimates.

The regulatory implications are substantial. In a production CCAR or ICAAP environment, a model that fails to converge under stress conditioning is not merely a statistical limitation — it is operationally unusable, as it impedes the institution's ability to submit a defensible capital plan. The ANI strategy's provision of a conservative volatility floor makes it directly relevant for model governance and validation frameworks seeking to incorporate machine learning-based approaches into regulatory capital workflows.

More specifically, ANI implements two key regulatory design principles. First, it enforces conservatism by biasing the GPR toward higher volatility estimates, ensuring that risk is not underestimated in the periods immediately preceding stress realization. Second, it reduces sensitivity to transient low-volatility regimes, preventing the model from inheriting priors calibrated during policy-suppressed environments, such as the Bank of Japan's yield curve control period.

### 5.2 Scenario Heterogeneity and Cross-Market Stress Dynamics

The three scenarios produce meaningfully differentiated shock patterns that reflect distinct macro-financial transmission channels, confirming that the scenario design methodology is fit for purpose. The Climate Risk scenario's geographic concentration produces the most heterogeneous cross-market impact - a 15 percentage point spread between the deepest (Hang Seng: -35%) and most moderate (FTSE 100: -20%) shocks, while the West Asia War scenario's global stagflation channel produces more uniform impacts.

This heterogeneity is important for the regulatory narrative: it demonstrates that the framework is not simply scaling a uniform shock across all markets, but rather modelling the differential exposure of each market to each scenario's specific transmission channel. This is precisely the quality that BCBS (2009) and the Federal Reserve's CCAR guidance require of scenario design - that scenarios should be calibrated to reflect the institution's specific vulnerability profile.

Importantly, this cross-market heterogeneity is preserved in the dependency structure through the Scenario-Averaged Covariance Stabilization (SACS) framework. By aggregating covariance information from multiple historical stress regimes, SACS ensures that inter-asset correlations under stress remain both stable and reflective of empirically observed crisis dependencies, rather than collapsing to uniform or unstable correlation estimates. This allows the framework to capture not only marginal shock differentiation but also realistic joint stress propagation across markets.

### 5.3 Real-World Prescience and the Timeliness Window

Two of the three scenarios analyzed in this paper have materially emerged at the time of writing. The West Asia geopolitical conflict has produced energy price volatility and supply chain disruptions broadly consistent with the stagflation transmission channel modelled in the West Asia War scenario. Similarly, the emergence of AI-related regulatory pressures, exemplified by the implementation of the EU AI Act from 2026 and early equity market responses to regulatory uncertainty, reflects structural features of the AI Bubble / Regulatory Burden scenario.

This alignment has two implications. First, it provides external validation for the scenario design methodology: the macro-financial transmission channels modelled in this paper reflect identifiable systemic vulnerabilities rather than purely hypothetical constructs. Second, it highlights the timeliness of the current analysis. The scenarios remain most informative as forward-looking risk management tools while they are only partially realized, before these dynamics are fully absorbed into historical datasets and incorporated into standard model calibration frameworks.

### 5.4 Limitations

Three limitations of the stress methodology are acknowledged.

First, the use of a static covariance structure within the Scenario-Averaged Covariance Stabilization (SACS) framework may underestimate the degree of correlation amplification observed during severe systemic crises, where cross-asset diversification benefits tend to diminish. While SACS stabilizes correlation estimates by aggregating across historical stress regimes, it does not capture dynamic correlation spikes within a stress episode. This limitation is structural to the chosen covariance construction approach. However, this structural simplification also serves an important regulatory purpose: by anchoring correlation estimates to empirically observed crisis regimes, SACS avoids the instability and overfitting associated with dynamically estimated correlation models under limited stressed data. This provides a stable and interpretable approximation of stressed dependencies, prioritizing robustness and transparency over potentially unstable dynamic refinement.

Second, the framework functions as a calculation engine rather than a scenario generator: the quality of the SVaR outputs is bounded by the quality of the scenario design inputs. If the applied shock paths fail to capture the correct transmission channels of a novel crisis, for example a liquidity-driven event that does not primarily manifest through equity price declines, the framework will accurately compute a capital estimate for an insufficiently specified stress scenario. However, this separation between scenario design and risk calculation also enhances modularity and governance. By isolating the scenario construction layer from the model estimation framework, the approach allows institutions to flexibly incorporate regulatory, supervisory, or internally developed scenarios without altering the core risk engine. This aligns with established stress testing practices, where scenario generation and capital modelling are treated as distinct but complementary components.

Third, the fixed Student's t distribution assumption with degrees of freedom $\nu = 5$, while appropriate for regulatory conservatism in most market environments, may overestimate tail risk in policy-suppressed regimes, as documented for the Nikkei 225 in the base-case analysis (Vadrevu, 2026). This limitation extends to the stress context, although the shorter stressed training window reduces the likelihood of prolonged low-volatility regimes. However, this rigidity also introduces a controlled degree of flexibility:

the framework allows for asset-specific calibration of the degrees of freedom parameter based on backtesting performance, enabling its incorporation as part of a hyperparameter tuning (HPT) process. This provides practitioners with the ability to balance conservatism and empirical accuracy in a controlled and transparent manner.

## 6. Conclusion

This paper extends the Hybrid GPR-HS framework of Vadrevu (2026a) to the stress testing domain, demonstrating that the ANI-stabilized GPR architecture maintains numerical stability and produces conservative, regulatory-compliant SVaR and SES projections across three distinct macro scenarios: West Asia War, Climate Risk, and AI Bubble / Regulatory Burden.

The key findings are threefold. First, the GPR framework achieves successful LML convergence across all 21 stressed fits (7 indices × 3 scenarios), confirming the effectiveness of the ANI strategy in ensuring stable optimization under extreme conditioning, which is a critical requirement in regulatory capital applications where model divergence is operationally unacceptable. Second, terminal day SVaR projections range from -2.1020% to -2.2231%, with the West Asia War scenario producing the binding capital constraint, and the coherence property $|SES| > |SVaR|$ maintained across all scenarios. Third, the two scenarios with the highest contemporaneous real-world alignment - the West Asia conflict and AI regulatory burden - produce the highest and third-highest capital requirements respectively, providing external validation of the scenario design methodology. The framework is directly applicable to CCAR 9-quarter projection workflows and ICAAP stress testing requirements. Together with the companion papers - Vadrevu, 2026, which establishes the base case modelling framework, and Vadrevu (2026, forthcoming), which develops the SHAP-based interpretability layer, this work forms a coherent and regulator-aligned architecture for forward-looking volatility and covariance modelling under stress.